\def\bq{\begin{equation}}
\def\eq{\end{equation}}
\def\bqa{\begin{eqnarray}}
\def\eqa{\end{eqnarray}}
\def\bqb{\begin{eqnarray*}}
\def\eqb{\end{eqnarray*}}
\documentstyle[12pt]{article}\textwidth 16cm
\def\pr#1#2#3{ Phys. Rev. ${\bf{#1}}$ (#2) #3}

\def\pl#1#2#3{ Phys. Lett. ${\bf{#1}}$ (#2) #3 }
\def\prep#1#2#3{ Phys. Re. ${\bf{#1}}$ (#2) #3}
\def\np#1#2#3{ Nucl. Phys. ${\bf{#1}}$ (#2) #3}
\def\zp#1#2#3{ Z. Phys. ${\bf{#1}}$ (#2) #3}
\def\ijmp#1#2#3{ Int. J. Mod. Phys. ${\bf{#1}}$ (#2) #3}

\def\ie{{\it i.e.\/}}
\def\eg{{\it e.g.\/}}

\global\nulldelimiterspace = 0pt

\def\Bsl{\hbox{/\kern-.6700em$B$}} %  Bslash
\def\Dsl{\hbox{/\kern-.6700em$D$}} %  Dslash
\def\Wsl{\hbox{/\kern-.6700em$W$}} %  Wslash

\def\roughly#1{\mathrel{\raise.3ex
    \hbox{$#1$\kern-.75em\lower1ex\hbox{$\sim$}}}}
\def\lsim{\roughly<}

\def\ol#1{\overline{#1}}

\def\L{ {\cal L }}
\def\O{ {\cal O }}

\def\mwd{M_W^2}

\def\lw{\lambda_W}

\begin{document}
\pagenumbering{arabic}
\thispagestyle{empty}
\def\thefootnote{\fnsymbol{footnote}}
\setcounter{footnote}{1}

\begin{flushright} CERN-TH/95-42 \\
 PM/95-04 \\ THES-TP 95/02 \\
February 1995 \\
 \end{flushright}
\vspace{2cm}
%---------------------titre ---------------------------------------
\begin{center}
{\Large\bf Anomalous Weak Boson Couplings: Suggestions from Unitarity
and Dynamics}
\footnote{Partially supported by the EC contract CHRX-CT94-0579.}
 \vspace{1.5cm}  \\
%-----------------------------------------------------------------
{\large G.J. Gounaris$^{a,b}$, F.M. Renard$^c$ and G.
Tsirigoti$^b$}
\vspace {0.5cm}  \\
$^a$Theory Division, CERN, \\
CH-1211 Geneva 23, Switzerland,\\
\vspace{0.2cm}
$^b$Department of Theoretical Physics, University of Thessaloniki,\\
Gr-54006, Thessaloniki, Greece,\\
\vspace{0.2cm}
$^c$Physique
Math\'{e}matique et Th\'{e}orique,
CNRS-URA 768,\\
Universit\'{e} de Montpellier II,
 F-34095 Montpellier Cedex 5.\\

\vspace {0.5cm}

{\bf Abstract}
\end{center}
\noindent
Taking into account the constraints from LEP1
and lower energy experiments,
we identify the seven $SU(2)\times U(1)$ gauge invariant
purely bosonic
$dim=6$ operators which provide a quite general ~description of how
New Physics could reflect in the bosonic world, if it happens
that all
new degrees of freedom are too heavy to be directly produced in the
future colliders. Five of these operators are CP conserving and the
remaining ones are CP violating. We derive the
unitarity constraints for the
CP violating operators and compare them with the already known
constraints for the CP conserving ones. Dynamical renormalizable
models are also presented, which partly elucidate what
the appearance of each of these operators can teach us on
the mechanism of spontaneous gauge symmetry breaking. \\
\vspace{0.5cm}
\begin{flushleft}
CERN-TH/95-42\\
February 95\\
\end{flushleft}
\def\thefootnote{\arabic{footnote}}
\setcounter{footnote}{0}
\clearpage

\section{Introduction}
The intense experimental effort to find  traces of New Physics
(NP) beyond the
Standard Model (SM) has so far given  only very  weak hints
\cite{lep1, lep1b}. No  new particles,
possibly associated with  NP
have ever been seen in our
present accelerators. Moreover,  the interactions
among the gauge bosons and the light fermions have been
thoroughly scrutinized at LEP1 and lower energies, and  were found
to be fully consistent with the
SM predictions. The only slight experimental hints for something
beyond the SM that exist at present
consist in the  well known  peculiarities observed
in $Z \to b \bar b$ \cite{lep1, lep1b}, and the persisting
indications favouring the  possibility of some
non-vanishing neutrino masses and a huge ~amount of dark matter
in the Universe.\par

Thus, before the excitation of new particles will
become possible, hopefully
in one of the contemplated future accelerators, it seems that our
main  hope to detect hints of NP is by carefully
searching for  anomalies
 in  interactions among the
gauge bosons, the Higgs and the quarks of the third family, since
these interactions have  not yet  been tested to the
same level of accuracy
as the light fermionic ones \cite{lep1, lep1b, GRVZbb}. Of course, if
the Higgs particle turns out to be above the TeV scale, it will itself
be part of  NP, inducing  new strong interactions mainly among the
longitudinal gauge bosons \cite{Chan}. Although this is a
viable possibility,
we  assume below that it will not be the case in Nature, and that the
Higgs will be discovered some day in the  mass range
of the electroweak
breaking scale $v=1/(\sqrt 2 G_\mu)^{1/2}$. \par

We therefore contemplate  an NP scenario according to which
the usual SM Higgs
particle exists and is, in a sense,  part of the ``old" physics.
 Moreover, in this  scenario
the NP scale $\Lambda_{NP}$, which determines the masses
of all new particles, is assumed to be very large.
 Under such conditions, a quite
general way of ~parametrizing NP is achieved by establishing
an effective Lagrangian
containing contributions from all possible
$SU(3)\times SU(2) \times U(1)$ gauge invariant operators
constructed from
scalar and gauge bosons fields, as well as the  quarks of the
third family (together with the  gluons). Since the contributions of
these  operators are  scaled
by inverse powers of $\Lambda_{NP}$,  it is plausible
to expect that for a sufficiently large NP
scale, the $dim=6$ operators
should give in general an adequate description \cite{Buchmuller}.
 Of course, it is quite possible that there exist NP aspects
whose scale is not really
very large, such as  the case of a moderately heavy vector
boson which mixes with $W$ or $Z$ \cite{BESS, Zprime}.
In such a case our approximation to retain only $dim=6$
operators might not be sufficient, and we would
have to include in our
expansion also higher dimensional operators. In the following we will
assume that this is not the case, though.\par

The complete list of  purely bosonic
 such $dim=6$ NP operators has been
known for some time\footnote{ These operators involve only weak
gauge bosons and Higgs.}  \cite{ Hag, HagCP},
 and recently we have also established the CP invariant
operators containing the $(t\ , b)$ quarks \cite{GRVZbb}.
In the present paper we focus on the purely bosonic operators though,
eleven of which conserve the  CP symmetry, while the
remaining five ones violate it.
We next take into account the fact that
seven of these purely bosonic
operators, (four CP conserving and three
violating ones) are already
excluded by  LEP1  and  low energy measurements
\cite{Hag, DeR, Bilal},
while another two
are completely insensitive to any conceivable experiments
\cite{Hag}.
Discarding all these irrelevant interactions,
we conclude that it should be sufficient
to describe the purely bosonic part of  NP below $\Lambda_{NP}$,
in terms of an effective Lagrangian which is
 a linear combination of the seven operators
called $\O_W$, $\O_{W\Phi}$, $\O_{B\Phi}$, $\O_{UW}$, $\O_{UB}$,
$\ol{\O}_{UW}$, $\ol{\O}_{UB}$. Presently existing LEP1 experiments
put only very moderate ~constraints on
these operators \cite{Hag, DeR}.
The general aim of the
present paper is to give an orientation on the magnitude of the
couplings of
these operators, based on considerations on the unitarity constraints
and on a class of dynamical scenarios.\par

The first three of these operators,
namely $\O_W$, $\O_{W\Phi}$ and  $\O_{B\Phi}$,
are the only ones  involving triple gauge boson couplings.
These operators are also CP symmetric, and are the only ones to
give anomalous contributions to the process
$e^+e^- \to W^+ W^-$, which will be studied at LEP2 and NLC
\cite{Bilenky, Layssac}.\par

The remaining four operators
only induce anomalous  $H\gamma \gamma$, $H\gamma Z$,  $H ZZ$
and $HWW$
couplings, and contain no triple gauge boson vertices
\cite{HagCP, Kramer}. The operators
$\O_{UW}$ and $\O_{UB}$  are CP symmetric,
while  $\ol{\O}_{UW}$, $\ol{\O}_{UB}$ are CP violating.
If $H$ is within the LEP2 range, these couplings could
be studied there by
carefully analysing Higgs-strahlung \cite{HagCP, Kramer, HZ}.
Thus, immediately after the ``hoped
for" discovery of the Higgs, the need to search for its
anomalous couplings
will arise.\par

In order to study the NP signatures  described by
the above operators,
it is very useful to first establish the unitarity constraints on their
couplings. Such unitarity constraints give  relations
between the strength
of these couplings and the energy scale where either unitarity will
be saturated, or (as  happened in the old Fermi theory) some of
the new degrees of freedom of NP will start being excited.
At the technical level such
relations are very helpful, since  they roughly
determine the energies
and couplings for which the  perturbative results are reliable.
In previous
works we have established the unitarity constraints for the five CP
conserving operators \cite{unitarity}. The first aim of the
present paper is to
complete this study by  giving the unitarity constraints also
for the CP violating
ones $\ol{\O}_{UW}$, $\ol{\O}_{UB}$. We then summarize
the implications
from the unitarity relations for all seven operators. These are the
``suggestions from Unitarity" alluded to in the title.\par

The second aim of the present work is to offer examples of dynamical
models containing new heavy degrees of freedom, which, after they are
integrated out, lead to an NP description in terms of  the purely
bosonic $dim=6$ operators. These examples are generalizations
of previous
ones given for the case of NP operators respecting custodial
$SU(2)_c$ symmetry \cite{dyn}.
The usefulness of such examples consists in
the fact that they provide a feeling on what type of anomalous
interactions
could be induced by various kinds of  new degrees of
freedom.  From these examples, we
infer that Higgs dependent operators
$\O_{UW}$, $\O_{UB}$, $\ol{\O}_{UW}$, $\ol{\O}_{UB}$, if they happen
to be created in the model, have their couplings determined by the
arbitrary
Yukawa type interactions in new physics. Thus, their couplings
are not constrained by existing experiments and have a
chance  to be observable at LEP2 \cite{HagCP}, \cite{HZ}.
On the contrary,
the purely gauge dependent operator $\O_W$ has its  strength
determined exclusively by the group properties of the NP  particles
we have integrated out. We would expect, therefore, that
$\O_W$ could  only become  appreciable if a
non-perturbative mechanism enhances it. The same is also true for
the operators $\O_{UW}$ and $\O_{UB}$, which were never generated
in these  models.\par

Disregarding all contributions which are either unobservable or are
very strongly constrained by existing experiments, and restricting to
$dim=6$ operators \cite{Hag, DeR, Bilal}, the
effective Lagrangian describing the purely bosonic part of NP
at the scale $\Lambda_{NP}$ is given by
\bqa
\L_{NP} & = & \lw \frac{g}{\mwd}\O_W +
{f_B g\prime \over {2M^2_W}}\O_{B\Phi} +{f_W g \over
{2M^2_W}}\O_{W\Phi}\ \ + \nonumber \\
\null & \null &
d\ \O_{UW} + {d_B\over4}\ \O_{UB} +\ol{d}\ \ol{\O}_{UW} +
{\ol{d}_B\over4}\ \ol{O}_{UB} \ \ \ \
\ \ ,  \
\eqa
where
\bqa
\O_{W\Phi} & = & i\, (D_\mu \Phi)^\dagger \overrightarrow \tau
\cdot \overrightarrow W^{\mu \nu} (D_\nu \Phi) \ \ \  , \ \
 \\[0.1cm]
\O_{B\Phi} & = & i\, (D_\mu \Phi)^\dagger B^{\mu \nu} (D_\nu
\Phi)\ \ \  , \\[0.1cm]
\O_W &= & {1\over3!}\left( \overrightarrow{W}^{\ \ \nu}_\mu\times
  \overrightarrow{W}^{\ \ \lambda}_\nu \right) \cdot
  \overrightarrow{W}^{\ \ \mu}_\lambda \ \ \
   \ \
\eqa
induce anomalous triple gauge boson couplings, while\footnote{In
the definition of $\O_{UW}$ and $\O_{UB}$ we have ~subtracted a
trivial contribution to the $W$ and $B$ kinetic energy respectively.}
\bqa
\O_{UW} & = & \frac{1}{v^2}\, (\Phi^\dagger \Phi - \frac{v^2}{2})
\, \overrightarrow W^{\mu\nu} \cdot \overrightarrow W_{\mu\nu} \ \
\  ,  \ \ \\[0.1cm]
\O_{UB} & = & \frac{4}{v^2}~ (\Phi^\dagger \Phi -\frac{v^2}{2})
B^{\mu\nu} \
B_{\mu\nu} \ \ \  , \ \ \\[0.1cm]
 \ol{\O}_{UW} & = & \frac{1}{v^2}\, (\Phi^\dagger
\Phi - \frac{v^2}{2})
\, \overrightarrow W^{\mu\nu} \cdot
\widetilde{\overrightarrow W}_{\mu\nu} \ \
\  ,  \ \ \\[0.1cm]
\ol{\O}_{UB} & = & \frac{4}{v^2}~ (\Phi^\dagger \Phi -\frac{v^2}{2})
B^{\mu\nu} \
\widetilde{B}_{\mu\nu} \ \ \  \ \ \
\eqa
create  anomalous CP conserving and CP violating
Higgs couplings. Note that
\bqa
\widetilde{B}_{\mu \nu}=\frac{1}{2}~\epsilon_{\mu \nu \rho \sigma}
B^{\rho \sigma} \ \ , \
\eqa
and similarly for $\widetilde{W}_{\mu \nu}$. \par

 Since the operators in (2)-(8) have a
dimension higher than four, they  will eventually
saturate  unitarity at
sufficiently high energies,
unless their locality is tempered by the excitation of new particles.
The unitarity constraints for
the CP conserving operators  shown in  (2)-(6)
have been derived in  \cite{unitarity}. They are given by
\bq
|f_B| \leq 98{M^2_W\over{s}} \ \ \ \ \ , \ \ \ \ \
 \ |f_W| \leq 31{M^2_W \over{s}} \ \ \ \ ,\
 \eq
\bq
|\lambda_W| \lsim 19~{M^2_W \over s} \ \ \  , \ \
\eq
\bq
 |d| \lsim 17.6~{M^2_W\over{s}}+2.43
{M_W\over{\sqrt{s}}} \ \ \ \  , \
\eq
\bq
-236~{M^2_W\over{s}}~+~1070~{M^3_W\over{s^{3/2}}}~ \lsim ~d_B ~\lsim
{}~ 192~{M^2_W\over{s}}~-~
1123~{M^3_W\over{s^{3/2}}}  \ \ \ \ ,\ \ \
\eq
where $s$ determines the square of the centre of mass energy of the
four-boson amplitude where unitarity is first reached.\par

Here we give the corresponding constraints
for the CP violating operators
$\ol{\O}_{UW}$ and $\ol{\O}_{UB}$. As in the previous cases, the most
important ones arise from the $j=0$ partial wave amplitudes
with vanishing total charge in the $s$-channel. For the
$\ol{\O}_{UB}$ case, the nine channels participating in the transition
matrix are  ($|\gamma\gamma\pm\pm \rangle$,
$|\gamma Z\pm\pm\rangle $,
$|ZZ\pm\pm \rangle $, $|ZZ LL \rangle $, $|W^-W^+LL \rangle $,
$|HH \rangle $), while for the $\ol{\O}_{UW}$ case
 there are  two additional channels given by
$|W^-W^+\pm \pm \rangle $. The whole procedure for $\ol{\O}_{UB}$ and
$\ol{\O}_{UW}$ is in
close analogy to the cases of the operators $\O_{UB}$ and $\O_{UW}$
treated in \cite{unitarity}, but this time the tree level
amplitudes are complex.
For the couplings defined in (1), we find
\bq
 |\ol{d}| \lsim 18.7~{M^2_W\over{s}}+3.04
{M_W\over{\sqrt{s}}} \ \ \ \  , \
\eq
\bq
 |\ol{d}_B| \lsim 176~{M^2_W\over{s}}-889
{M_W\over{\sqrt{s}}} \ \ \ \  . \
\eq
Applying (10)-(15) for $s=1\, TeV^2$, we get $|f_B|\lsim 0.6$,
$|f_W| \lsim 0.2$, $|\lw | \lsim 0.12$, $ |d| \lsim 0.3 $,
$ |d_B| \lsim 0.7 $, $ |\ol{d}| \lsim 0.3 $, $ |\ol{d}_B| \lsim 0.7 $.
There are two remarks  to be made concerning these relations.
The first is
that the constraints for the CP conserving and the CP violating
Higgs interactions,
derived from  ($\O_{UW}$, $\O_{UB}$) and ($\ol{\O}_{UW}$,
$\ol{\O}_{UB}$)
respectively, are quite similar to each other. The second
remark is that the unitarity constraints for the
$\overrightarrow W_{\mu \nu}$ involving operators
($\O_{W\Phi}$, $\O_W$, $\O_{UW}$, $\ol{\O}_{UW}$),
are a factor of 2 to 3
stronger than the ~corresponding ones for the $B_{\mu \nu}$
involving operators. This means that for similar  NP couplings,
the new physics forces in the
$WW$  channel are  considerably stronger than the
forces  in the $ZZ$ one. A similar  situation is known
to be valid also
for the SM interactions.\par

\vspace{0.2cm}
To get a feeling of what kind of NP couplings one might expect to
appear in (1), we now turn to specific dynamical models. The only
requirement in  these models is that they always
respect $SU(2)\times U(1)$ gauge symmetry and renormalizability.
As in the usual SM Lagrangian, no  additional discrete
symmetry like \eg\@  CP invariance is imposed.\par

\vspace{0.2cm}
\noindent
\underline{Model A:}\par
In this model, we assume that NP is determined by a complex scalar
field $\Psi$, which  has isospin $I$ and  hypercharge $Y$
under the $SU(2)\times U(1)$ gauge group. Since  $\Psi$
acquires its mass before the electroweak spontaneous breaking,
this mass must be large,\@ \ie\@   $M \equiv \Lambda_{NP} \gg v$.
The $\Psi$ may also have a   hyper-colour
$\tilde{N}_c$. The basic renormalizable Lagrangian will then
be the sum of the usual SM Lagrangian $\L_{SM}$ and the Lagrangian
\bq
\L_{\psi}= D_\mu \Psi^\dagger\  D^\mu \Psi
 - \Lambda_{NP}^2\Psi^\dagger \Psi +
2 g_{\psi 1} (\Psi^\dagger \Psi)(\Phi^\dagger \Phi) +
g_{\psi 2} \left [ (\Psi^\dagger \widetilde \Phi)
(\widetilde \Phi^\dagger  \Psi) -
(\Psi^\dagger  \Phi) ( \Phi^\dagger  \Psi)\right ] \  ,
\eq
describing the interactions of the $\Psi$ field.
In writing (16)  we have omitted irrelevant $(\Psi^\dagger \Psi)^2$
terms and  we have also taken
$I\not= 0$ and $( Y \not= 1/6,\ 7/6,\ -5/6\ ,... )$,
so that to exclude
a direct ($\Psi$ - $\Phi$) mixing  and a possible coupling
of a single
$\Psi $ with either the scalar or the fermion fields of the SM.\par

The standard techniques  may now be used
to obtain the effective Lagrangian describing the ~electroweak
interactions at a scale just below $\Lambda_{NP}$. This is achieved by
integrating out, at the one-loop order, the heavy field  $\Psi$.
Thus, by employing the Seeley--DeWitt expansion of the relevant
determinant, we obtain the following NP contribution to the
electroweak interactions
at this scale:
\bqa
\L_{NP} & = &\frac{(2I+1) \widetilde{N}_c}{(4\pi)^2}~
\Bigg\{-~2g_{\psi 1} \Lambda^2_{NP} \left (\frac{1}{\epsilon}+1
\right )(\Phi^\dagger \Phi)
{}~+~ \frac{2}{ \epsilon}\left (g^2_{\psi 1} +g^2_{\psi 2}~
\frac{I(I+1)}{3}\right ) (\Phi^\dagger \Phi)^2
 \nonumber\\[0.2cm]
\null & \null &
- ~\frac{1}{12}\ \left (
\frac{1}{\epsilon} +\frac{g_{\psi 1} v^2}{\Lambda^2_{NP}} \right )
\left [
\frac{g^2 I(I+1)}{3}\ \overrightarrow W_{\mu
\nu} \overrightarrow W^{\mu \nu}
+  Y^2 g\prime ^2 \ B_{\mu \nu}B^{\mu \nu} \right ]
\nonumber \\[0.2cm]
\null & \null &
+ ~ \frac{8}{6 \Lambda^2_{NP}}(g^3_{\psi 1} + g_{\psi 1}
g^2_{\psi 2}I(I+1)) (\Phi^\dagger \Phi) ^3 ~
 \nonumber\\[0.2cm]
\null & \null &
+ ~ \frac{1}{3 \Lambda^2_{NP}}\
\left (g^2_{\psi 1} +g^2_{\psi 2}\frac{I(I+1)}{3}\right )
\partial_\mu (\Phi^\dagger \Phi)\partial^\mu (\Phi^\dagger \Phi)
\nonumber \\[0.2cm]
\null & \null &
{}~+~ g^2_{\psi 2}\frac{4 I(I+1)}{9\Lambda^2_{NP}} \left
[ (\Phi^\dagger \Phi)
(D_\mu \Phi^\dagger D^\mu \Phi) - (D_\mu \Phi^\dagger \Phi)
(\Phi^\dagger D^\mu \Phi) \right ]  \nonumber\\[0.3cm]
\null & \null &
- ~ \frac{g^2 I(I+1)}{90\Lambda^2_{NP}}\left [
  g \O_W + \frac{1}{4}\, \overline{\O}_{DW} +
 5 v^2 g_{\psi 1} \O_{UW} \right]  \nonumber \\[0.2cm]
\null & \null &
+~\frac{g\prime Y}{\Lambda^2_{NP}} \left [
 \frac{2 I(I+1)}{9}\ g g_{\psi 2} \O_{BW} -g\prime g_{\psi 1} Y
\frac{v^2}{24} \O_{UB} -\frac{g\prime Y}{120} \O_{DB} \right ]
\Bigg\} \ \ , \
\eqa
where $\epsilon=2-n/2$ (with $n$  the number of dimensions)
is the usual dimensional regularization
parameter, and (4)-(6) are used together with the definitions
\bqa
\overline{\O}_{DW} & =& 2~ (D_{\mu} \overrightarrow W^{\mu
\rho}) (D^{\nu} \overrightarrow W_{\nu \rho})  \ \ \
  , \ \  \\[0.1cm]
\O_{DB} & = & (\partial_{\mu}B_{\nu \rho})(\partial^\mu B^{\nu
\rho}) \ \ \  , \ \ \\[0.1cm]
\O_{BW} & =& \frac{1}{2}~ \Phi^\dagger B_{\mu \nu}
\overrightarrow \tau \cdot \overrightarrow W^{\mu \nu} \Phi
\ \ \  . \
\eqa\par

The first three terms  in $\L_{NP}$ just renormalize scalar and gauge
boson terms already   existing
in $\L_{SM}$, while the next two   indicate an example
of how the NP
can generate the two  unobservable operators
$(\Phi ^\dagger \Phi )^3$ and
$\partial_\mu (\Phi^\dagger \Phi)\partial^\mu (\Phi^\dagger \Phi)$
mentioned in the introduction. The operator
$(\Phi^\dagger \Phi)
(D_\mu \Phi^\dagger D^\mu \Phi) - (D_\mu \Phi^\dagger \Phi)
(\Phi^\dagger D^\mu \Phi)$, as well as the $\ol{\O}_{DW}$,
$\O_{DB}$, $\O_{BW}$, given in (18)-(20), should be negligible,
according to LEP1 experiments. For reasonable $\Psi$ isospin
and hypercharge,
this is easily understood for the operators $\ol{\O}_{DW}$ and
$\O_{DB}$, whose couplings are proportional to $g^2$ and $g\prime ^2$
respectively. The negligible strength of the other two operators
just mentioned can  be ~accommodated if we assume that
$g_{\psi 2}$ (defined in  (16)) is negligible.\par

The interesting thing about Model A is that there is nothing that
prohibits  $g_{\psi 1}$ (also
defined in  (16)) to be large. And if this does happen,  then  the
operators $\O_{UW}$ and $\O_{UB}$  will be  proportionally
enhanced by  NP. Unfortunately,  an analogous   enhancement for
$\O_W$  is not so easy. The  $\O_W$ coupling expected perturbatively
satisfies $\lw \sim g^2$, and   it
should therefore be  similar to the coupling of the strongly
constrained operator $\ol{\O}_{DW}$ (see (18)).
Only if $\O_W$ is somehow  ~non-perturbatively enhanced with respect
to  $\ol{\O}_{DW}$, by a mechanism like the one discussed in \cite{dyn},
we could hope that it would become observable. \par

Therefore, out of the seven operators appearing in (1), Model A
favours only $\O_{UW}$ and $\O_{UB}$, and to a lesser extent $\O_W$.
The couplings of the Higgs involving operators $\O_{UW}$ and $\O_{UB}$
depend on the unknown physics of
the scalar sector. Thus, these two later
operators really teach us something about the mechanism that breaks
spontaneously the gauge symmetry. On the contrary
the purely gauge
dependent operator $\O_W$ seems naturally suppressed
at the perturbative
level, by the small coupling $g$. Nevertheless,
it is at least generated in
this model. Note that if $\Psi$ had $Y=0$, then only the custodially
$SU(2)_c$ invariant operators $\O_{UW}$ and $\O_W$ would have
appeared\footnote{ Provided of course that we still keep
the assumption
$g_{\psi 2}\sim 0$.}
\cite{dyn}.

\vspace{0.2cm}
\noindent
\underline{Model B:}\par

We now turn to Model B where NP is determined instead  by a fermion
field $F$ whose left and right component have the same isospin $I$ and
hypercharge $Y$. Because of the vectorial character of the
model, there are no
anomalies  and  $F$ acquires its mass before the spontaneous
electroweak breaking takes place.
Hence,  we can assume that $F$  has a very large mass
$\Lambda_{NP}$ and possibly also a hyper-colour $\tilde{N}_c$.
To construct the basic renormalizable Lagrangian we should now add to
$\L_{SM}$ the term
\bq
\L_F=~i \ol{F} (\rlap/\partial +i g\,{\overrightarrow
\Wsl} \cdot \overrightarrow{t} +i g\prime  Y \Bsl )F
-\Lambda_{NP} \ol{F}F \  \ ,  \ \
\eq
with $\overrightarrow{t}$ denoting the isospin $I$ matrices.
In writing (21) we have excluded a discrete set of hypercharge and
isospin values which would allow a coupling  of $F$ with the SM
fermions and possibly also with the standard Higgs. \par

Integrating the fermion loop as before \cite{Ball},
we get  at the scale
$\Lambda_{NP}$:
\bqa
\L_{NP} & = & \frac{(2I+1) \widetilde{N}_c}{ (4\pi)^2}~
\Bigg \{ -\ \frac{g^2 I(I+1)}{9\epsilon }\ \overrightarrow W_{\mu
\nu} \overrightarrow W^{\mu \nu}
\ -\  \frac{Y^2 g\prime ^2}{3 \epsilon}  \ B_{\mu \nu}B^{\mu \nu}
\nonumber \\
\null & \null &
+\ \frac{g^2 I(I+1)}{45 \Lambda^2_{NP}}\left [
  g \O_W - \, \ol{\O}_{DW}
\right]~-~ \frac{g\prime ^2 Y^2}{15 \Lambda^2_{NP}}\
\O_{DB} \Bigg \}
\ \ . \
\eqa\par

The only interesting operator generated in this case is $\O_W$,
about which though (as well as about the unwanted operators
$\ol{\O}_{DW}$ and $\O_{DB}$), the same remarks apply as
in Model A.
It seems that if NP only includes fermionic new degrees of freedom,
we cannot learn much about the scalar sector by studying the
anomalous bosonic couplings. To reiterate on this we thus turn to
Model B$^\prime$.\par

%\vspace{0.2cm}
\noindent
\underline{Model B$^\prime$ :}\par

To the preceding NP spectrum we just add a heavy
($M_s \gg v$) scalar
 field $S^0$ with vanishing isospin and hypercharge.
Then the most general renormalizable interaction to be added to
$\L_{SM}$ becomes
\bqa
\L_{FS}& = & i\ol{F}\Dsl F -\Lambda_{NP}\ol{F}F +
g_f S^0\ol{F}F +
  i {g_f}\prime S^0 \ol{F} \gamma_5 F  \nonumber\\
\null & \null & g_\phi M_s S^0 \Phi^\dagger \Phi
 + \frac{1}{2}(\partial S)^2 -\frac{M_s^2}{2}(S^0)^2 \ \ , \
\eqa
where  irrelevant ${(S^0)}^3$ and ${(S^0)}^4$ terms have been omitted.
Integrating out first the heavy $F$ field, and then substituting
$S^0$ to $g_\phi \Phi^\dagger \Phi/M_S$, we find that NP  generates,
in addition to the terms appearing in (22), contributions also from
all four  purely Higgs operators shown in (1), \ie\@
$\O_{UW}$, $\O_{UB}$, $\ol{\O}_{UW}$, $\ol{\O}_{UB}$. Restricting for
simplicity to  $I=1/2$ for the isospin of the $F$ fermion, the
couplings of the CP conserving operators are expressed as
\bqa
d & = & -\left (\frac{g^2 v^2 \widetilde{N}_c}
     {48 \pi^2 \Lambda_{NP} M_s} \right ) \     g_f g_\phi \ \ \ , \ \\
d_B & = & -\left (\frac{g\prime ^2 v^2 Y^2 \widetilde{N}_c}
     {48 \pi^2 \Lambda_{NP} M_s} \right ) \     g_f  g_\phi \ \ \ , \
\eqa
while those of the CP violating ones satisfy
\bq
\frac {\ol{d}}{d}~=~ \frac{\ol{d}_B}{d_B}~=~-\frac{{g_f}\prime}{g_f}\ \
\ . \ \
\eq
Note that for vanishing hypercharge $Y$ for the $F$ fermion,
only the custodially $SU(2)_c$
invariant operators $O_{UW}$ and $\ol{O}_{UW}$ would be generated
\cite{dyn}. Since the $g_f$ and ${g_f}\prime$ couplings in (23) are
a priori on the same footing, we conclude that in this model all
four operators $\O_{UW}$, $\O_{UB}$, $\ol{\O}_{UW}$, $\ol{\O}_{UB}$
can be generated with appreciable couplings. We also note that
if the scalar boson $S$ were chosen instead to be isovector, then the
model would have generated the operator $\O_{BW}$ and its CP violating
analogue. Since the last two operators are very ~strongly
constrained from existing experiments, we would conclude that
such a situation is ~disfavoured.\par

The above considerations lead to the conclusion
that the question whether  one of the operators
($\O_{UW}$, $\O_{UB}$, $\ol{\O}_{UW}$, $\ol{\O}_{UB}$)
will be generated or not is
intimately connected with the nature of the mechanism responsible
for the spontaneous breaking of the gauge symmetry.
 The experimental search for such an operator will  teach us
something on how the spontaneous symmetry breaking
works. Our models imply also that,
to a lesser extent, $\O_W$ can also be generated by NP; but this
operator seems to be rather independent of the scalar sector. Finally
the other two operators in (1), namely $\O_{W \Phi}$ and $\O_{B\Phi}$,
were never generated in our models.\par

It is unnecessary to state that we take these models
only as indicative. The
laws of New Physics are certainly much more elaborate than our
toy models suggest. It could also be that the residual
interactions below the NP scale $\Lambda_{NP}$ not only involve weak
bosons but also heavy quarks, \ie\@ the third family. In \cite{GRVZbb} we
have established the list of 28 $dim=6$ operators involving the third
family, 14 of them involving the $t_R$ field (which in SM is
associated to the
top mass generation) and we showed that
some of them could also be at the origin of the
departure of $Zb\bar b$ from SM predictions.\par
 In any case an active experimental search
at LEP2 and at higher energy colliders should be made in order
to identify any of the  operators we have discussed. Once any
of them is
found, then (as in the old Fermi theory), the unitarity
constraints presented above may    help
deciding how far we  are from the energy region where some new
degrees of freedom should start being excited.\par

\vspace{0.5cm}
\underline{Acknowledgements}: We would like to thank Jacques Layssac
for his help in the derivation of the numerical expressions of
the unitarity bounds.\par

\end{document}